# THE CONNECTION AMONG GAMMA-RAY BURST HOST-GALAXIES, BL LACS AND QUASARS[1]

Anup Rej
Department of Physics (Section Lade), NTNU, 7034 Trondheim, Norway
*anup.rej@phys.ntnu.no*

15th January 1999

---

[1]The paper is available in the web in the address: http://www.phys.ntnu.no/~rej

**Abstract**

A possible connection among host-galaxies of gamma-ray bursts, BL Lacs and quasars is analysed. It is believed that the gamma-ray bursts, which do not show radio or infrared emission, occur in faint blue dwarf galaxies, that are seen around radio-quiet quasars, which lie in clusters. The GRB counterparts, which show radio emission, may be associated with more evolved starbursting environments, and arise from red galaxies, that are observed around some radio-loud quasars lying in richer clusters. The absorption lines are believed to arise from gases, that move in the intergalactic medium, due to tidal interactions occuring among the compact galaxies in the cluster environment. The connections between intra-day variable BL Lac-blazars and radio emitting gamma-ray hosts are also studied. The hosts of gamma-ray bursts, BL Lacs and quasars are believed to represent different evolutionary phases of a particular type of galaxy with a "twisted" morphology. They mostly occur in star formation environments and evolve from a galaxy possessing no AGN, but consisting of very massive young stellar sytems, to galaxies possessing developed AGNs, like in quasars, whose gas content have been exhausted in giving birth to the stars at the AGN core. In between these two phases, these galaxies may pass through the state of the red objects, which contain a new born quasar hidden under dust.This evolutionary history of the morphologies and the environment, where GRBs may occur is believed to be related to the process of formation of galaxies and large scale structures. Moreover, the starbursting peaks at different redshifts may indicate a universe that is different from the standard cosmolgy.

# 1. Introduction

After the discoveries of several optical counterparts the gamma-ray bursts are clearly established to be extragalactic phenomena, which occur mostly at high-redshifts. We (Rej & Østgaard 1997) proposed earlier that the gamma-ray bursts could arise due to "Failed-Type I supernpva/hypernova" in the star forming regions and pointed out that the two peaks, which are observed in the distribution of the BATSE counts, when plotted against the duration of the bursts, may correspond to the two epochs when the star formation peak. The corresponding redshifts are around $z \sim 1$ and $z \sim 3$.

Though the redshift measurements of the four optical counterparts support these ideas, the counterparts of GRB 980425 fall outside the above regions. However, this gamma-ray burst has been observed at the same place where the supernova SN 1998bw has been detected, which strengthens the idea of the "failed supernova and hypernova" scenario.

In a recent paper (Rej 1998a) the relations among red quasars, BL Lacs, and Blazars with gamma-ray bursts are discussed in the light of this "supernova/hypernova" scenario. It is proposed that in the starburst environments, where SN Type I and Type II occur, the star formation will cause outflows of alpha element enriched material, that will give rise to MgII absorption lines, that are present in the spectra of GRB 970508 and GRB 980703.

In the 19th Texas Symposium in Paris I have reported (Rej 1998b) the possibility of the relations among the gamma-ray burst host-galaxies, BL Lacs and quasars. I have suggested that the cluster environments, where the quasars are observed, could be the possible sites to look for the host-galaxies of the gamma-ray bursts. The possibility of ESO 184-G82, enigmatic NGC 1275, and QSO 1229+240 being different stages of evolution in the formation of quasars, BL Lacs and blazars is also pointed out. It is believed that the host-galaxies, in which the gamma-ray bursts occur, may be the progenitor of the galaxies that host quasars, BL Lacs and blazars while the systems evolve to develop active AGNs. In such GRB host-galaxies one should expect to observe blue super-clusters of stars as they are seen in NGC 1275, for example. As indicated (Rej 1998b), the host-galaxies, where the gamma-ray bursts occur, are believed to possess homogeneous morphologies, that I have called "compact twisted galaxies", which undergo intense star formation. With the star formation these systems evolve from blue starburst galaxies to infrared luminous sytems harbouring nascent quasars hidden under dust, before developing active AGNs, like in quasars.

In this paper I shall summarize the ideas, that I have reported earlier, and seek to build an understanding about the nature of the host-galaxies where the gamma-ray bursts take place and illucidate their connections with quasars and BL Lacs.

## 2. Observational indications

*2.1 GRB 970228*
- Gamma-ray bursts are extragalactic
- The surrounding nebulosity belongs to a host galaxy that is unusually blue
- Reddening in the beginning could be due to absorption by the intervening gas through which the ejecta passes
- The proper motion observed could be interpreted as arising due to relativistically moving plasma moving along a "jet"

*2.2. GRB 970508*
- The presence of faint blue galaxy nearby suggests active star formation region
- The absorption line features at z=0.835, and a weak MgII absorbing system at z=0.768, together with the compactness of the region, make it a good candidate for a star forming region
- The weak MgII absorption system bears a great resemblance to BL Lac PKS 0138-097
- The radio flares indicate shock induced plasma emission as in intra-day variable QSOs

*2.3. GRB 971214*
- No strong emission line in the beginning. However, two weeks later a slightly extended emission feature, that resembles BL Lacs
- Broad absorption features
- The host galaxy was found to be at redshift 3.43

*2.4. GRB 980329*
- Very faint at visible wavelengths, brighter at infrared, and still brighter at radio wavelengths
- Intra day variable radio source, like in GRB 970508
- Relatively bright sub-millimeter flux indicating a dusty star forming region
- The radio through sub-millimeter spectrum is well fitted with a power-law with index 0.9

*2.5. GRB 980425*
- The "peculiar supernova" SN 1998bw could be due to core-collapse of very massive Wolf-Rayet star, that may eject relativistic jet in the process of the collapse

*2.6. GRB 980703*
- A compact starburst object corresponding to a blue host galaxy
- The strong emisson line and absorption lines, identified as Fe II and Mg II lines at redshift 0.97, superposed on an otherwise featureless UV continuum
- No broad emission lines are seen and consistent with a young compact unobsucred starburst galaxy

## 3. Questions

It is so far believed that the host-galaxies, where the GRB counterparts have been found, resemble spirals, or barred spirals. However, a closer inspection reveals that they could be very unlike these known galaxies. Instead they may well belong to a morphology, which resembles more the "irregular galaxies". In fact, they may have a twisted morphology, like coiled columns of gas forming knots, where star formations occur. Systems, which look similar, are seen in association with quasars and they are interpreted as arising due to interactions of a galaxy with its close companion, believed to be in the process of undergoing merging events. This brings into the question: Are the gamma-ray bursts related to the environments of the quasars? Are the star formation, the quasar formation, and the gamma-ray bursts all related to each other?

Visually it appears that the nearby galaxy ESO 184 G-82, where GRB 980425 and SN 1998bw have been discovered, possesses highly coiled columns of gas, that twist to form "knots" where super star clusters may one day form. However, such systems do not show any evolved AGN possessing an accretion disk around. This brings forth the question: Are the host-galaxies, where gamma-ray bursts occur, related to the host galaxies in which the quasars will be born one day? Will these twisted faint compact galaxies evolve to become active AGNs? One may also be inclined to wonder: Do quasars, BL Lacs and blazars represent different stages of evolution of such galaxies - only possessing more evolved and active AGNs, and older stellar population while they evolve?

In fact, there is an evidence of correlation between gamma-ray bursts and radio-quiet quasars at a confidence level of more than 99% (Schartel et al. 1997). However, there is no evidence for correlation between GRBs, seyferts, radio-loud quasars, or BL Lacs. One may wonder if it may have something to do with the evolutionary history of the GRB host-galaxies as well as the environments where they appear.

In this connection it should be noted that a significant excess of blue galaxies and emission-line candidates, spatially associated with the quasars, have been observed. The quasar sample at z ~ 1.1 suggests that the QSOs are in compact groups, or cluster of star-forming galaxies (Hutchings et al. 1994). These faint galaxies are irregular and compact in size. At lower redshifts ( < 0.5) quasars reside in small to moderate clusters (Yee & Green 1984; Haymann 1990; Smith & Heckman 1990), while at higher redshifts a marked difference in the environment of radio-loud and radio-quiet QSOs (Yee & Green 1987; Yee & Ellingson 1993) exists. Radio-loud QSOs are associated with rich clusters (Hall & Green 1998), while the radio-quiet quasars appear in smaller groups.

The presence of quasars (Bagla 1997) in the cluster at high-redshift indicates relations between quasars and the galaxy formation. In fact, quasars could be significantly related to the population of Lyman alpha galaxies, which show substantial clustering at z ~ 3.1 (Cristiani 1998) and interpreted as progenitors of the massive galxies of the present epoch, or percursor of the present day clusters (Governato et al. 1998). Hutchings et al. (1998) have noted that tidal events are more marked cause (or effect) of QSO activities at higher redshifts, than the interactions at lower regime. These observations have implications for galaxy formation, and cluster evolution in relation to quasars. One may wonder if the clusters, where the radio-quiet quasars appear, and the clusters, where the radio-loud quasars are observed, represent differnet evolutionary phases of the similar groups of starbursting galaxies. Do the radio-quiet quasars evolve into radio-loud objects as the cluster of galaxies become more densely packed ?

Highly significant evidence for the existence of large clusters of galaxies at a redshift ~3.1 (Stiedel et al. 1998) have been found. The clustering structure of these star-forming galaxies is , in many way, similar to a less evolved version of the structure found at z=0.985 (Le Fevre et al. 1994). One may be inclined to believe that these structure formations may correspond to the redshift epochs where the star formations peak. The spectra of the starburst galaxies at high redshifts show many similarities with those of the nearby starbursts and demonstrate that by z>3 massive galaxy formation was well under way (Giavalisco et al. 1998).

Or does such association of the quasars with clusters of galaxies indicate an evolution and formation of galaxies and clusters, that are very different from what is presently believed? H. C. Arp (1997) has argued that quasars are ejected from active galactic centres. These ejected quasars start out with low luminosities and high z. As they travel away from the galaxy of origin, they evolve and grow in size and decay in redshift. At certain quantized values of the redshifts the quasars become highly active and break up into many objects which finally evolve into groups and clusters? Such scenario will demand episodic creation in a non-expanding universe. If so, our present understanding of the cosmos would be in serious jeopardy.

**4. Gamma-ray bursts and quasars**

*4.1.Optical Transient (OT) and QSO*
The first indication of a relation between QSOs and GRBs came with *GB 910219,* which was detected by WATCH aboard the Soviet Space Observatory GRANAT . It consisted of two main bright peaks, separated by about 50 s with a duration of 10 s each. The estimated fluence was 10-5 ergs cm-2 between 6 and 120 keV. The error box of localization of GB 910219 was related to the OT error ellipse that included a QSO with z=1.78.

*4.2. GRBs and radio-quiet quasars*
While one has not found any clear correlation with clusters of galaxies, or the super-galactic plane, which exclude the possibility that any normal consituents of galaxies may cause the gamma-ray bursts, Schartel et al. (1997) have reported evidence for a positional correlation between gamma-ray bursts and radio-quiet quasars. However, their studies have been restricted to bursts with the smallest error box. As the position errors of GRBs are functions of the gamma-flux, the results are biased towards the brightest bursts, which are nearer. They argued that only intrinsically brighter radio-quiet quasars are correlated with such GRBs.

Hurley et al.(1998) re-examined the above claim of the correlation between the gamma-ray bursts and the radio-quiet quasars using larger sample of gamma-ray bursts and more precise location of the bursts. Their error boxes had areas much smaller than the BATSE error circles. Using these reduced error boxes they found no correlation at all. This negative result indicated that gamma-ray bursts do not arise from quasars.

*4.3. Quasar environments*
However, this observation does not exclude the possibility that the gamma-ray bursts occur in the faint galaxies that are found in the environments around the quasars.

Jager et al. (1998) obtained deep R-band images (R ~24.5) of 20 quasars in the range 0.75<z<0.85. Sixteen of the quasars were radio-quiet. In almost all cases close faint companions were found. In half of the cases a trend of increasing number counts towards the QSOs were noted. In 25% of the cases an excess appears more compact and concentrated within <200 kpc around the QSOs, suggesting that they reside in groups of galaxies.

At redshift 1.1, in the environments of the quasars, the distribution of the emission line candidates is found to be similar to the blue objects suggesting the environment of active star formation.

The redshift distribution of 1 Jy quasar sample also peaks at z ~ 1 while the quasar samples, derived from radio surveys with lower flux density limits, peak close to the redshift peak which is found for the optically selected quasars.

*4.4. Interactions*
HST studies of the quasars at lower redshift ( <0.3) show interactions (Bahcall et al. 1997). About 65% of the sample studied optically show close companions. Three in the sample of 20, show intense interactions. Interestingly, two of these three quasars are radio-quiet. In all three cases the tidal arms are seen.

Hubble space telescope images of z~1 also reveal diversity of morphologies including small compact galaxies, and apparently normal late type spirals, and give evidence of a high frequency of merging events, and interactions.

It is certainly possible that some of the faint blue companion galaxies, that are detected in association with the quasars, in fact, undergo star formation while they interact in cluster and evolve to form larger structures. As results of supernovae explosions these dwarf galaxies blow out metal rich wind.

However, due to an observed correlation of the absorption cross section with the identified galaxy K-luminosity, it is difficult to reconcile a large metal contribution from the supernovae alone (Churchill et al. 1996). In fact, observations of nearby galaxies show that the characteristics of Mg II absorption lines depend on a wide variety of mechanisms. The large velocity range, which the absorption covers, is reminiscent of absorption lines from the High Velocity Clouds (Bowen et al. 1996). One of the explanations of such HVCs is that they result from the interactions between the Galaxy and the neighbouring satellites. Moreover, many MgII clouds have ionization structures in which C IV surrounds their lower ionization Ly-alpha -MgII cores ( Churchill 1998). This may also be the case for interacting galaxies that one observes in association with quasars. It suggests that the galactic mass has connection with a galaxy's ability to organize the tidally stripped, accreting or infalling material. The systems, identified with MgII absorption, could indeed be associated with the tidal processes.

The study of Hutchings et al. ( 1998) suggests that the tidal events are more marked in QSO activities at higher redshifts, than the observations at lower redshifts.

### 4.5. Absorption lines

Indeed absorption lines are found in all high redshift QSOs. At least four classes of line systems exist. Type A consist of very broad absorption troughs with z's different from $z_{em}$ by upto 0.1 c. Type Bs are the ones with sharp lines having absorption redshifts not very different from the emission redshifts. Types Cs and Ds are those with sharp lines with $(z_{em} - z_{abs})/(1+z_{em}) > 0.01$. C-system contain metal lines, but the D-system consist of larger number of Lyman absorption lines seen at wavelengths below the Lya emission.

The great majority of absorption lines arise outside the emission lines region. However, many absorption line regions can be very close to QSOs. Perhaps they arise from clouds in, or near, galaxies in a cluster that also contains the QSOs.

## 5. Mg II absorption systems

### 5.1. General

In fact, a relatively luminous galaxy, within ~40kpc of the QSO line-of-sight, appears to be a prerequisite for the detection of MgII absorption. It has been found that the spatial distribution of the absorbing gas surrounding intermediate redshift galaxies is not smoothly varying, and the velocities of the gas clouds cannot be described by a single systematic kinematic model. Absorption profiles arising in 0.4< z<1.0 galaxies, in fact, exhibit a rich variety of sub-component structure and kinematic complexity.

### 5.2. Characteristics of the absorbing galaxy

Typical MgII absorbing galaxy, at intermediate redshifts, is characterized by an inner region of radius ~ 15 kpc, that gives rise to damped Ly-alpha lines, together with a region extending to ~ 40 kpc, that produces MgII absorption lines and Lyman limit break, and an outer region extending to ~ 70 kpc that is less shielded from extragalactic background radiation and produces absorption lines in higher ions, such as CIV.

### 5.3. LSB/ dwarf galaxies

At lower redshifts, the general population of low surface brightness galaxies outnumber high surface brightness galaxies by a factor of at least two, and the Ly-alpha clouds are more directly associated with low surface brightness and/or dwarf galaxies, or with the remnant material left over from the formation of galaxies and/or small galaxy groups. The discovery of saturated MgII doublet, associated with the z=0.072 dwarf galaxy, suggests that the star formation in these objects may directly govern their gas cross section. If so, active star formation dwarf galaxies contribute to the overall metal line absorption cross section. One would then expect that the abundance pattern arising from a bursting dwarf to be alpha-group enhanced.

During the star formation the Type II SNe disperse alpha-group elements such as silicon and magnesium, over a short time. Following the star formation peak, the chemical build up of iron-group elements ( a SN Type Ia process) gradually increases, and the Fe/Mg abundance ratio rises. If the alpha group elements are enhanced relative to Fe group, then the chemical enrichment is believed to be dominated by Type II SNe. If a given Ly-alpha cloud is measured to have Fe/H >-1 and alpha/Fe group abundance ratios approach solar properties, then one might infer that the Type Ia SNe have played a role. In a star-forming region, the SN Type II rate is believed to be about one order of magnitude larger than the SN Type Ia rate, as observed for late type spirals. In this case the FeII emission is produced mainly by shocks from SN Type II and is a measure of the star formation activity in the recent past.

The example of the HIRES/Keck spectrum of PKS 0454+039 for MgII absorption in Ly-alpha clouds show two clouds at z=0.6248 and z=0.9315. The cloud at z=.93 has super-solar metallicity 0.5<Z/Z-solar<1.5. There are a few diffuse extended LSB galaxies over the impact parameter 20- 60 kpc. The two clouds are

## 6. Star formations and quasars

There are also indications that the quasar epoch may indeed be the epoch of the star formation (Shaver et al. 1998). A striking similarity exists between the redshift evolution of the space density of flat-spectrum QSOs and the evolution of the star formation rate (SFR). Moreover, the evolution of the radio luminosity density of the luminous radio sources is similar to the evolution of the UV luminosity density of star forming galaxies. Even at redshifts higher than z=1 the agreement between QSOs and star forming galaxy samples, both in the location of the maximum and the high redshift decay rate, is found. The space density of the radio-selected quasars shows a turn-over (Hook et al. 1998) at high redshift ( z>3). The decline in space density at high redshifts has a similar form to what is observed for the optically-selected quasars. The observed luminosity function (LF) of QSOs and its redshift evolution can be explained with a starburst model. Furthermore, the sample of luminous starburst galaxies gives evidence for the presence of two or more nuclei with bridges and tails, that are hallmarks of tidally interacting systems. Superwinds may carry substantial amount of metal out of the starbursts. Though typical MgII absorption line systems may be due to tidal interactions.

## 7. Galaxy formation

From the observations of galaxy formation of blocks in rich groups in the redshift range 2.6 <z<3.9, Clements and Couch (1996) concluded that this range could be the epoch when the galaxies are formed through mergers. Pascarelle et al. (1995) observed subgalactic blocks at z=2.39 and discussed their relation to galaxy formation. The discovery of the groups involving faint blue galaxies in the environment of the quasars also imply connection between the quasar formation epoch and the epoch of the galaxy formation. The transitional mass from a large galaxy to a small group is around 5x $10^{12}$ M-solar. In hierarchical cosmologies this corresponds to z = 3 ± 0.5, where star formation peak. Besides that, the class of irregular/ merger galaxies, which are relatively rare at bright magnitudes, make up about a half, or a third of all galaxies with $I_{AB}$ ~ 25. The median redshift at this apparent magnitude is z ~ 0.8. This lies around the other redshift region where the star formations show the second redshift peak.

At z~1 intense star formation activities occur in about half of the MgII high redshift absorber subsample. Photometric redshifts for galaxies brighter than $I_{ST}$=28 in the Hubble Deep Field show two peaks, that can be understood if larger galaxies form stars early at z ~ 3, and if star formation is delayed in the dwarf systems until z ~ 1. It is believed that the collapse of the dwarf galaxy-scale masses could be inhibited until z~ 1 due to phtoionization of the IGM. These dwarfs will fade after a star formation at 0.5< z <1.0.

The peak of the bright quasars also occurs at the star formation peak, and the evolution of the quasars at high redshifts can be closely related with the formation of the galaxies in overdense regions (Haenelt & Rees 1993). Ly-alpha emitting objects have been found around z ~ 2.5 QSOs (Pascarelle et al. 1996, Warren & Møller 1996) supporting the above view. The study of the very high redshift radio-quiet quasar BR 1202-0725 at z=4.7 (McMahon et.al 1994) also shows star formation in its environment (Fontana et al. 1996). The observed spectral properties of its companion and the velocity profile of the Ly-alpha emitter indicate tidal interactions (Fontana et al. 1998). The experimental evidences indicate a high redshift galaxy, which is still far from a dynamic equilibrium and undergoing high star formation in its individual "pre-galactic" components.

## 8. Quasars and clustering

Shaver (1984) found possible clustering of quasars in the Veron Catalog. Several authors have observed (Andreani & Cristiani 1992; Mo & Fang 1993; Croom & Shanks 1996) slightly larger clustering than what is observed in the present day galaxies ( but lower than the clustering of the clusters). Recently La Franca et al. (1998) have found an evidence for an increase in clustering with increasing redshift for QSOs with 0.3 < z <2.2. Thus the most luminous objects in the high redshift universe seem to exhibit strong spatial clustering.

At least there are two known cases (Cristiani 1998 and references therein), which show large quasar groups at z<2.0: One is at z ~1.1 and the other around z ~1.9. They show clustering similar to present-day galaxies. However, Stephens et al. (1997) have reported evidence for an increase in clustering towards high redshift. For redshifts > 2.7 they have found much higher clustering of the quasars compared to present-day galaxies.

## 9. Extreme Red Objects (EROs) and red quasars

Though an excess of small compact faint objects have been found around high-redshifts quasars, and radio galaxies (Dressler et al 1993,1994; Hutchings 1995; Pascarelle et al. 1996), a suprising number of extremely red objects (EROs), too, have been identified in the vicinity of high redshift radio galaxies and quasars (McCarthy et al. 1995; Eisenhardt & Dickinson 1992; Hu & Ridgeqay 1994; Graham et al. 1994; Dey et al. 1995; Soifer et al. 1994). They are optically extremely faint (R > 24). These objects are spatially extended and a few appear to be weak radio sources. One of the possible explanations of the extreme red colour could be redenning caused by starburst. For example, Yamada et al. (1997) have found many fairly luminous red galaxies around radio-loud quasar 1335.8+2834. A clustering of very red objects at z=1.1 is also observed. Their optical-NIR colours and magnitudes are consistent with brighter cluster galaxies. There are several emission line galaxies near the aforesaid quasar. Another example of clustering of red galaxies near radio-loud object 3C 324 has been found by Dickinson (Dickinson 1995; 1997). Hintzen et al. (1991) found clustering of galaxies, which are more luminous in R-band than the brightest galaxies in the present day clusters, around z=0.9 -1.5 radio-loud quasars.

Graham & Dey (1996) have made infrared observations of one of the reddest EROs, H10. They inferred that the assymetric morphology of H10, whose line luminosity is in accord with that of a Seyfert II, is consistent with that of an interacting galaxy at z=1.44. The presence of emission line also suggests star formation activities and the system seems to be dusty. In the local universe such galaxies are represented in the sample of IRAS bright galaxies (Soifer et al. 1987).

Recently Francis et al. (1998) have reported a small number of radio-quiet high redshift galaxies, which share the red clours of the high redshift radio galaxies. For example the galaxy 2139-4434 B1 at redshift 2.38, is associated with a cluster of QSO absorption-line systems, and two other radio-quiet galaxies nearby are also such red galaxies.

The flat-spectrum radio quasars are found to be red quasars, whose reddening could be due to dust obscuration. Selecting the reddest of the sample one finds that about one quarter of them are BL Lacs. The red quasars have faint, extended, galaxy-like emission in the optical, with little or no evidence for bright point source, while NIR images reveal a dominant point source in most cases.

## 10. BL Lacertae objects

*10.1. General*
Great similarity between BL Lac spectrum of PKS 0138-097 and the optical spectrum of GRB 970508, and the abundance of BL Lacs amongst known bright gamma-ray sources have been suspected to give a positive correlation. However, the morphology of the host-galaxies and the environment of the BL Lacs are poorly known with respect to the quasars. Many BL Lacs are often described as bright point sources in underlying galaxies. The overall morphology of host-galaxies is believed to be little perturbed, and different from radio-loud quasars (Falomo et al. 1998). The preliminary investigation of the close environment of BL Lacs has suggested that asymmetries and/or distortion of the nebulous surrounding could be a rather frequent phenomenon (Falomo, Melnik & Tanzi 1990). They possess low luminosity extended radio structures, that often resemble FR I galaxies with weak radio luminosity. However, many of the radio selected BL Lacs have an overall energy distribution similar to the flat radio-spectrum quasars, with weak and broad emission lines, and extended radio structures that are more luminous than FR Is.

Radio spectra of BL Lacs remain optically thick upto about 90 GHz (QSOs are optically thin above 10 GHz). About two-thirds of the flux come from within a region of 0.3-0.5 mas, and an elongated structure of 1.5 X 0.5 mas.

The host galaxies are believed to be somewhat fainter than brightest cluster galaxies (Heidt 1998). On average they are fainter than a typical FR I host, and resemble rather a FR II host. Virtually all BL Lacs show signatures of absorbing systems. Faint companions have been detected for a substantial number of BL Lacs, and evidence for interactions with bright companions have been found. The close companions seem to occur rather frequently and seem to indicate "merging" events (Falomo 1996). Very similar companions are often seen within a few arcsec from quasars (cf. Bahcall et al. 1995;Disney et al. 1995). The lumnosities of the close BL Lac companions are also similar to the close companions of QSOs. However, it is believed that BL Lacs rest in poorer cluster environment. Both the host galaxies and environment of BL Lacs indicate evolution with redshift. The host becomes brighter as the environment becomes denser.

The current sample of the BL Lacs is divided into "red" and "blue" types (Urry et al. 1998). The "red" BL Lacs have peak energy output at infrared-optical region, while the "blue" BL Lacs peak at UV-X-ray wavelenghts. In the sample studied, in the redshift range 0.19 - 1.0, the "blue" objects have a systematic lower redshift than the "red" objects. However, Urry et al. have found no difference in morphologies between the "red" and the "blue" types.

Tran et al. (1998) have indicated an evolutionary connection of ultralumionus infrared galaxies(ULIRGs) with quasars. They have found evidence that some ULIGRs are powered by quasars, and some are energized by starbursts with no AGN hidden under dust. They inferred that the buried quasars preferentially reside in ULGIRs, that have warm colours and possess spectra characteristics of high-ionization Seyfert 2 galaxies. This may suggest that young quasar is born while the system undergoes starbursts and the young AGN remains hidden under dust. The stellar population should be older in such systems.

*10.2. MgII absorber (Stocke & Rector 1997)*
A number of MgII systems in the 1Jy radio-selected BL Lacs sample have been found. Some of the absorbers could be intrinsic to BL Lacs, or there may be a correlation of absorbing gas in the foreground, and the nearly featureless spectra.

For example, BL Lac 0454+844 shows Mg II doublet at z=1.34. It is the most distant known in the 1 Jy sample. BL Lac 2029+121 has emission lines of C IV, C III and Mg II at z=1.215 with foreground Mg II, Mg I , Fe II/Mn 2600, 2606 Å and Fe II 2382Å absorption at z =1.117. BL Lac 0138-097 too possesses Mg II absorption doublet. Emission lines of weak Mg II and O II correspond to a redshift of z=0.733.

*10.3. BL Lac blazar AO 0235+164*
In the light of the above discussion it is interesting to study the BL Lac object AO 0235+164. It consists of at least three spectral line systems: One weak emission line system of Mg II l2800, [Ne V] l3426 and [O II] l3727 at redshift of 0.94, and two absorption line systems at redshifts of 0.54 and 0.85. The z=0.94 system shows hydrogen lines $H_d$ and $H_g$. The equivalent width of the Hg line, which is 7.2Å in the rest frame of AO 0235+164, exceeds the 5 Å limit for an object to be included in the 1 Jy BL Lac sample of Stickel et al. (1991). The z=0.94 system shows broad permitted Mg II and narrow forbidden lines. The spectrum is comparable to a small number of other BL Lacs with nuclear line emission. The line intensities vary over a time scale of a few years (Cohen et al. 1987). The faint nebulous extension, which is a "companion" galaxy, has narrow emission lines of [O II], [O III] and Hb at redshift 0.542 with R magnitude of 20.9. But the host galaxy of the system at redshift 0.94 has remained undetected. The "companion" appears to possess two curved plumes, or arms on the opposite side of a core. It is an AGN surrounded by faint nebulosity, and shows spectra that have broad absorption lines displaced to the short-wavelength side of C IV, Si IV and N V emission lines, like some of the less extreme BALQSOs. The emission lines of Lyα, C IV l1549, and C III l1909 have broad emission wings, characteristic of a QSO. The Lyα emission has a sharp central spike, but no broad absorption. Moreover the C IV is asymmetric: The red (low velocity) side is steeper than the blue, and the blue edge of the trough shows possible structure. The C IV absorption extends from roughly 1500 to 6000 km/s. The spectrum of the "companion" is similar to the spectra of the majority of BALQSO, and the strong absorption features are all high ionization metal lines. It is a very active object ejecting $N^{4+}$, $C^{3+}$, $Si^{3+}$ at high velocities into the intergalactic medium. The Seyfert 1s and the narrow emission line galaxies with [O III]/Hb < 3 and emission line FWHM < 300 km/s, have [O II] luminosity 5-10 times smaller than the total [O II] luminosity observed at the "companion" and the intervening galaxies at z=0.525. The closest match of the "companion" galaxy is Markarian 833. There are several extremely faint ( R~ 25) pairs of objects near AO 0235+164 which could be part of z=0.851 or z=0.94 group (Nilsson et al. 1996).

## 11. Discussion

From the above discussions and the analysis that I have made in a previous paper (Rej 1998a) the connections between the starburst galaxies and the gamma-ray bursts appear abundantly clear. The three counterparts for which the measurements of redshifts are available indicate that the host galaxies of the GRBs lie in the redshifts where the star formation peaks. The unusually blue colours of the host galaxies of GRB 970228, GRB 980703 , and the presence of a faint blue galaxy near GRB 970508, and the correlations found with quasars and GRBs, all indicate that the gamma-ray bursts have taken place in the faint galaxies around quasars. The absorption lines clearly indicate situations that are observed in the surrounding of quasars, which show the presence of the tidal interactions among the companion galaxies in the cluster giving rise to MgII absorption systems. The magnitude of the optical counterparts are also in conformity with what are observed in the faint galaxies around quasars. The reddening of the gamma-burst, as in the case of GRB 970228 is reminiscent of the presence of obscuring dust in the starburst environment. The presence of a brighter radio and infrared luminosity compared to the very faint optical visibility in the case of GRB 980329 also supports the dust obscuration mechanism in starburst galaxies. The spectrum of the GRB 970508 resembles a BL lac, and the appearance of emission lines two week later in an otherwise featureless spectrum of GRB 971214 give evidence for its connection with a blazar. When faint, the optical spectra of a blazar show emission lines like normal QSOs, but during outburst the emission lines are washed away as in a BL Lac. The variability in the radio and mm domain of the GRB 980329 indicates that it could well be a BL lac, like the variable radio source found in association with GRB 970508. There is a dust reddening in GRB 980329, which could possibly indicate an ERO object, or a "red" BL lac. The VLBI observations of GRB 970508 show that the radio source is unresolved(<0.3 mas). During the first month the 8.46 GHz and 4.86 GHz flux densities of this GRB showed rapid fluctuations (Frail et al. 1997). Bremer et al. (1998) found a maximum around 12 days at 86 GHz. A shallow maximum was observed at 8.46 GHz near 55 days (Frail et al. 1997). In a highly variable blazar, like AO 0235+164, the radio variations have been found to be correlated over at least a factor of 50 frequency range. The low frequency variations are found to be an extension of the high frequency variations and intrinsic to the source itself. The shortest rise and fall time scale is found to be of weeks duration for GHz frequencies, and months for the lower frequencies for AO 0235+164. This behaviour continues in the mm domain. Similar situation probably prevails in the case of GRB 970508. No detection at 232 GHz for GRB 970508 after 14 days is consistent with this scenario. Thus there is evidence that supports the belief that a few of the GRBs have occured in BL Lac blazars with high variability. Non detection of radio flux at 8.46 GHz about a day after the burst in the case of GRB 971214 does not exclude the variability in the radio range in a larger frequency domain over months. Therefore its associtaion with a BL Lac blazar is not excluded.

Then comes the question of the lack of correlation between the GRBs and the BL Lacs. If the objects involved in the GRBs, where radio emissions are detected, resemble more like AO 0235+164 blazar, it could be the cause for the non detection of a correlation. In fact, Schartel et al. have found correlation at 95% level between GRBs and a group of objects that they have called AG (Active Galaxies). This group typically consists of nuclear radio, or X-ray source. It is possible that GRB hosts having radio emsission are associated with these objects, which are grouped under AG. Additionally ultra-luminous infrared galaxies, that are difficult to detect in the optical, might be a second possibility.

Following these arguments it will be worth while to investigate the correlations between blazars, ultra-luminous infra-red galaxies and GRBs. As the nature of the clustering, as well as the galaxies in the environment could play important roles in the case of the radio-quiet quasars, it will be more desirable to study the correlations including larger error boxes, and including more fainter bursts. Particularly the correlations between the "Active Galaxies" and the GRBs should be carefully investigated.

However, to understand the phenomenon of the gamma-ray bursts fully, one needs to understand the relations among star formation, quasar formation and galaxy formation in the universe. In this perspective a better comprehension of the nature of the host galaxies of the quasars and the BL Lacs and their environments will be required. Especially the evolutionary dynamics of these objects, and their environments, should be of special interest. Moreover an understanding about the nature of clustering of galaxies, and their evolution with redshifts, would be highly valuable before one may be able to put the pieces of the puzzle about the gamma-ray bursts together.

Besides what is said above, the theories of the star formation and the galaxy formation may require a revision. One may have to abandon the standard paradigm of the black-holes and the accretion disks feeding the AGNs in explaining the dynamics of quasars. It is well possible that a very different mechanism might be at play than what is believed today. Especially the phenomenon known as "merging" companions in relation to quasars and BL Lacs should be revisited. According to the scenario, that I have sketched, they may represent evolutionary dynamics of the way the quasars are formed, and to which the host-galaxies of the gamma-ray bursts may belong at the nascent stage. To understand this dynamics one will need to study closely the morphologies of the hosts of gamma-ray bursts, quasars and BL Lacs, in particular. This may in turn pave a way in understanding how the galaxies and the large-scale structures are formed in the universe.

Given the observational evidences in different wavelenghts, that have been made in the last few years, it is not at all convincing that the standard theory about the origin of the universe is on a sound footing. In this connection it is extremely interesting to observe the two redshift regions when star formation, quasar formation and galaxy formation seem to peak. Instead of following the wisdom of the General Relativity, which predicts a big-bang universe, the universe may have been created through episodic bursts of creations that occur at particular "quantized" values of redshifts( Arp et al. 1990,1992; Arp 1997). Do galaxies spew matter out and break into smaller units, before assembling again into larger structures? The galaxies similar to the present day universe ( and may be even larger) seem to be present at very high-redshifts. Between the very high redshift universe and the present epoch there seem to exist regions where the universe is dominated by smaller and bluer galaxies undergoing intense star formations. They may thus be sowing the seeds of the formations of larger structures. In such a scenario one will be required to abandon the standard cosmology.

**12. Concluding remarks about the gamma-ray bursts**

*i) GRBs associated with quasars and BL Lacs*
The MgII systems, detected in GRB 970508, and GRB 980703 indicate environments of interacting galaxies in star-forming regions at high-redshift as they are in Quasars and BL Lacs.

*ii) GRBs occur in host-galaxies lying in the environment of interacting galaxies or so-called "mergers*
The observations of the counterparts of gamma-ray bursts GRB 970228, and GRB 980425 indicate that the bursts occur offset from the centre of the host galaxies. These galaxies show elongated and twisted features without an evolved AGN and are of actively starburst type. Such faint galaxies are observed in the environment of quasars lying in clusters of compact galaxies. We believe that the gamma-ray bursts have occurred while "hypernova" has exploded in the region where massive stellar systems are in the process of formation. Though the mechanism of this "hypernova" explosion is still an enigma.

The counterparts of gamma-ray burst GRB 970508 and GRB 980329 show variable radio emissions as in the environment of some blazars. They could be associated with such BL Lac balazars with morphologies, which are observed in the BL Lacs accompanied by "merger" galaxies. These radio emitting hosts could possibly be in a more evolved state compared to the radio-quiet ones. They could be associated with compact red objects EROs, found around radio-loud quasars at high redshifts. Such host galaxies will show diffuse nebulosities that extend to larger areas, and are irregularly distributed. The central radio source will remain unresolved, as in BL Lacs, and the gamma-ray burst will occur in a more central region compared to GRB 970228. One may find a nascent quasar hidden under dust causing reddening as in ULIRGs.

*iii) GRBs occur in the star-forming regions*
Luminous galaxies placed at redshift $z \sim 3$, or dwarfs around $z \sim 1$, will show the magnitude range as has been observed for the counterparts. We conclude that the majority of the gamma-ray bursts will occur in the redshift regions where the star formation peak. It could well be possible that the star formation rate around the higher-redshift peak is substantially higher than the star formation rate around the lower-redshift peak.

*iv) The radio output is caused by ejecta which cause the evolution from high-energy to low-energy emission features*
The observations indicate that the evolution of the emissions starting from the high-energy to the low-frequency radio band could be caused by ejecta, as in the case of blazars. When the relativistically moving plasma slows down, and the sub-relativistic plasma catches it up from behind, long-lasting X-ray afterglow can be produced in this scenario.

*v) Supernova*
Although the supernovae decay at slower rate in the visible wavelengths, than what are observed with the optical counterparts, they can decay much faster in UV. Thus a supernova at high-redshift could appear to decay in the visual, which is UV in the rest frame, at the observed rate. The fast decline of the counterparts indicate that they could be supernovae decaying in UV at high frequencies.

*The final conclusion:* The GRBs are generated by collapse of very massive stellar objects in starburst galaxies, like the faint blue dwarf companions, which are seen around radio-quiet quasars. Some may also arise in red galaxies, that accompany radio-loud quasars and seen in association with BL Lac blazars and infra-red galaxies, where new born quasars may remain hidden under dust. The gamma-ray light curves can be explained by invoking inverse Compton scattering of soft photons from blobs of relativistically moving plasma ejected in the process of the collapse. The mechanisms of such collapse and ejection of blobs should be investigated.

## 13. Other speculations and predictions

- The host galaxies, where GRB's occur belong to the same morphology, that does not fit into the known classification scheme of the galaxies.
- The GRB's will occur in these "twisted galaxies" that have not yet formed any central core or much nuclear activities.
- These twisted galaxies may be the progenitor galaxies that host QSOs, BL Lacs, and Blazars.
- The GRB host-galaxies, that have a loose structure like twisting knots, evolve with the star formation activities and wind up in a tighter form, that resemble a kind of spiral where most of the gases are exhausted in the arms. At that stage these galaxies host QSOs.

- At the less evolved stage such galaxies will resemble, for example, the enigmatic galaxy NGC 1275, that is a strong radio source and shows fetaures of "off the beam" BL Lacertae-like object.
- In the GRB host galaxies one should definitely find super blue clusters of stars as in NGC 1275.
- The so-called interacting systems and mergers, where QSOs are observed, probably represent different stages of evolution of a particular irregular type- galaxies.
- The "twisted" galaxies are bound systems, from where clouds cannot escape easily except through the ejection of blobs from the nuclei, which grow by feeding on the gases in the twisted arms.
- There should exist correlations among the QSOs, GRBs and intensely star forming blue galaxies with no/ or little developed nuclei.
- A completely different mechanism than the model where black-holes are involved will be required to explain the phenomena of QSOs.
- It may well be possible that our understanding of the universe is wrong.

*Acknowledgments :* I should like to express my gratitude to Professor Hans Kolbenstvedt and Professor Erlend Østgaard for extending a cordial and helpful relation with the Department of Physics, NTNU(Lade).